
\font\twelverm=cmr10 scaled 1200    \font\twelvei=cmmi10 scaled 1200
\font\twelvesy=cmsy10 scaled 1200   \font\twelveex=cmex10 scaled 1200
\font\twelvebf=cmbx10 scaled 1200   \font\twelvesl=cmsl10 scaled 1200
\font\twelvett=cmtt10 scaled 1200   \font\twelveit=cmti10 scaled 1200
\skewchar\twelvei='177   \skewchar\twelvesy='60
\def\twelvepoint{\normalbaselineskip=12.4pt
  \abovedisplayskip 12.4pt plus 3pt minus 9pt
  \belowdisplayskip 12.4pt plus 3pt minus 9pt
\abovedisplayshortskip 0pt plus 3pt \belowdisplayshortskip 7.2pt plus
3pt minus 4pt
  \smallskipamount=3.6pt plus1.2pt minus1.2pt
  \medskipamount=7.2pt plus2.4pt minus2.4pt
  \bigskipamount=14.4pt plus4.8pt minus4.8pt
  \def\rm{\fam0\twelverm}          \def\it{\fam\itfam\twelveit}%
  \def\sl{\fam\slfam\twelvesl}     \def\bf{\fam\bffam\twelvebf}%
  \def\mit{\fam 1}                 \def\cal{\fam 2}%
  \def\tt{\twelvett}
  \textfont0=\twelverm   \scriptfont0=\tenrm   \scriptscriptfont0=\sevenrm
  \textfont1=\twelvei    \scriptfont1=\teni    \scriptscriptfont1=\seveni
  \textfont2=\twelvesy   \scriptfont2=\tensy   \scriptscriptfont2=\sevensy
  \textfont3=\twelveex   \scriptfont3=\twelveex  \scriptscriptfont3=\twelveex
  \textfont\itfam=\twelveit
  \textfont\slfam=\twelvesl
  \textfont\bffam=\twelvebf \scriptfont\bffam=\tenbf
  \scriptscriptfont\bffam=\sevenbf
  \normalbaselines\rm}

\def\beginlinemode{\endmode
  \begingroup\parskip=0pt \obeylines\def\\{\par}\def\endmode{\par\endgroup}}
\def\beginparmode{\endmode
  \begingroup \def\endmode{\par\endgroup}}
\let\endmode=\par
{\obeylines\gdef\
{}}
\def\singlespace{\baselineskip=\normalbaselineskip}
\def\oneandahalfspace{\baselineskip=\normalbaselineskip
  \multiply\baselineskip by 3 \divide\baselineskip by 2}
\def\doublespace{\baselineskip=\normalbaselineskip \multiply\baselineskip by 2}
\newcount\firstpageno
\firstpageno=2
\footline={\ifnum\pageno<\firstpageno{\hfil}\else{\hfil
\twelverm\folio\hfil}\fi}
\let\rawfootnote=\footnote              
\def\footnote#1#2{{\rm\singlespace\parindent=0pt\rawfootnote{#1}{#2}}}
\def\raggedcenter{\leftskip=2em plus 12em \rightskip=\leftskip
  \parindent=0pt \parfillskip=0pt \spaceskip=.3333em \xspaceskip=.5em
  \pretolerance=9999 \tolerance=9999
  \hyphenpenalty=9999 \exhyphenpenalty=9999 }
\parskip=\medskipamount
\twelvepoint            
\overfullrule=0pt       
\def\preprintno#1{
 \rightline{\rm #1}}    
\def\author                     
  {\vskip 3pt plus 0.2fill \beginlinemode
   \singlespace \raggedcenter \twelvesc}
\def\affil                      
  {\vskip 3pt plus 0.1fill \beginlinemode
   \oneandahalfspace \raggedcenter \sl}
\def\abstract                   
  {\vskip 3pt plus 0.3fill \beginparmode
   \doublespace \narrower \noindent ABSTRACT: }
\def\endtitlepage               
  {\endpage                     
   \body}
\def\body                       
  {\beginparmode}               

\def\subhead#1{                 
  \vskip 0.25truein             
  {\raggedcenter #1 \par}
   \nobreak\vskip 0.1truein\nobreak}
\def\refto#1{$|{#1}$}           
\def\references                 
  {\subhead{References}         
   \beginparmode
   \frenchspacing \parindent=0pt \leftskip=1truecm
   \parskip=8pt plus 3pt \everypar{\hangindent=\parindent}}
\gdef\refis#1{\indent\hbox to 0pt{\hss#1.~}}    
\gdef\journal#1, #2, #3, 1#4#5#6{               
    {\sl #1~}{\bf #2}, #3, (1#4#5#6)}           
\def\refstylenp{                
  \gdef\refto##1{ [##1]}                                
  \gdef\refis##1{\indent\hbox to 0pt{\hss##1)~}}        
  \gdef\journal##1, ##2, ##3, ##4 {                     
     {\sl ##1~}{\bf ##2~}(##3) ##4 }}
\def\refstyleprnp{              
  \gdef\refto##1{ [##1]}                                
  \gdef\refis##1{\indent\hbox to 0pt{\hss##1)~}}        
  \gdef\journal##1, ##2, ##3, 1##4##5##6{               
    {\sl ##1~}{\bf ##2~}(1##4##5##6) ##3}}
\def\pr{\journal Phys. Rev., }

\def\prl{\journal Phys. Rev. Lett., }
\def\prpts{\journal Phys. Rep., }
\def\np{\journal Nucl. Phys., }
\def\pl{\journal Phys. Lett., }

\def\endreferences{\body}
\def\endpage                    
  {\vfill\eject}
\def\endpaper                   
  {\endmode\vfill\supereject}
\def\endit
  {\endpaper\end}
\def\ref#1{Ref. #1}                     
\def\Ref#1{Ref. #1}                     

\def\m@th{\mathsurround=0pt }
\font\twelvesc=cmcsc10 scaled 1200
\def\cite#1{{#1}}
\def\(#1){(\call{#1})}
\def\call#1{{#1}}
\def\taghead#1{}
\def\leaderfill{\leaders\hbox to 1em{\hss.\hss}\hfill}
\def\twiddle{\lower.9ex\rlap{$\kern-.1em\scriptstyle\sim$}}
\def\bigtwiddle{\lower1.ex\rlap{$\sim$}}
\def\gtwid{\mathrel{\raise.3ex\hbox{$>$\kern-.75em\lower1ex\hbox{$\sim$}}}}
\def\ltwid{\mathrel{\raise.3ex\hbox{$<$\kern-.75em\lower1ex\hbox{$\sim$}}}}
\def\square{\kern1pt\vbox{\hrule height 1.2pt\hbox{\vrule width 1.2pt\hskip 3pt
   \vbox{\vskip 6pt}\hskip 3pt\vrule width 0.6pt}\hrule height 0.6pt}\kern1pt}
\catcode`@=11
\newcount\tagnumber\tagnumber=0

\immediate\newwrite\eqnfile
\newif\if@qnfile\@qnfilefalse
\def\write@qn#1{}
\def\writenew@qn#1{}
\def\w@rnwrite#1{\write@qn{#1}\message{#1}}
\def\@rrwrite#1{\write@qn{#1}\errmessage{#1}}

\def\taghead#1{\gdef\t@ghead{#1}\global\tagnumber=0}
\def\t@ghead{}

\expandafter\def\csname @qnnum-3\endcsname
  {{\t@ghead\advance\tagnumber by -3\relax\number\tagnumber}}
\expandafter\def\csname @qnnum-2\endcsname
  {{\t@ghead\advance\tagnumber by -2\relax\number\tagnumber}}
\expandafter\def\csname @qnnum-1\endcsname
  {{\t@ghead\advance\tagnumber by -1\relax\number\tagnumber}}
\expandafter\def\csname @qnnum0\endcsname
  {\t@ghead\number\tagnumber}
\expandafter\def\csname @qnnum+1\endcsname
  {{\t@ghead\advance\tagnumber by 1\relax\number\tagnumber}}
\expandafter\def\csname @qnnum+2\endcsname
  {{\t@ghead\advance\tagnumber by 2\relax\number\tagnumber}}
\expandafter\def\csname @qnnum+3\endcsname
  {{\t@ghead\advance\tagnumber by 3\relax\number\tagnumber}}

\def\equationfile{%
  \@qnfiletrue\immediate\openout\eqnfile=\jobname.eqn%
  \def\write@qn##1{\if@qnfile\immediate\write\eqnfile{##1}\fi}
  \def\writenew@qn##1{\if@qnfile\immediate\write\eqnfile
    {\noexpand\tag{##1} = (\t@ghead\number\tagnumber)}\fi}
}

\def\callall#1{\xdef#1##1{#1{\noexpand\call{##1}}}}
\def\call#1{\each@rg\callr@nge{#1}}

\def\each@rg#1#2{{\let\thecsname=#1\expandafter\first@rg#2,\end,}}
\def\first@rg#1,{\thecsname{#1}\apply@rg}
\def\apply@rg#1,{\ifx\end#1\let\next=\relax%
\else,\thecsname{#1}\let\next=\apply@rg\fi\next}

\def\callr@nge#1{\calldor@nge#1-\end-}
\def\callr@ngeat#1\end-{#1}
\def\calldor@nge#1-#2-{\ifx\end#2\@qneatspace#1 %
  \else\calll@@p{#1}{#2}\callr@ngeat\fi}
\def\calll@@p#1#2{\ifnum#1>#2{\@rrwrite{Equation range #1-#2\space is bad.}
\errhelp{If you call a series of equations by the notation M-N, then M and
N must be integers, and N must be greater than or equal to M.}}\else%
 {\count0=#1\count1=#2\advance\count1
by1\relax\expandafter\@qncall\the\count0,%
  \loop\advance\count0 by1\relax%
    \ifnum\count0<\count1,\expandafter\@qncall\the\count0,%
  \repeat}\fi}

\def\@qneatspace#1#2 {\@qncall#1#2,}
\def\@qncall#1,{\ifunc@lled{#1}{\def\next{#1}\ifx\next\empty\else
  \w@rnwrite{Equation number \noexpand\(>>#1<<) has not been defined yet.}
  >>#1<<\fi}\else\csname @qnnum#1\endcsname\fi}

\let\eqnono=\eqno
\def\eqno(#1){\tag#1}
\def\tag#1$${\eqnono(\displayt@g#1 )$$}

\def\aligntag#1\endaligntag
  $${\gdef\tag##1\\{&(##1 )\cr}\eqalignno{#1\\}$$
  \gdef\tag##1$${\eqnono(\displayt@g##1 )$$}}

\def\eqalignno#1{\displ@y \tabskip\centering
  \halign to\displaywidth{\hfil$\displaystyle{##}$\tabskip\z@skip
    &$\displaystyle{{}##}$\hfil\tabskip\centering
    &\llap{$\displayt@gpar##$}\tabskip\z@skip\crcr
    #1\crcr}}

\def\displayt@gpar(#1){(\displayt@g#1 )}

\def\displayt@g#1 {\rm\ifunc@lled{#1}\global\advance\tagnumber by1
        {\def\next{#1}\ifx\next\empty\else\expandafter
        \xdef\csname @qnnum#1\endcsname{\t@ghead\number\tagnumber}\fi}%
  \writenew@qn{#1}\t@ghead\number\tagnumber\else
        {\edef\next{\t@ghead\number\tagnumber}%
        \expandafter\ifx\csname @qnnum#1\endcsname\next\else
        \w@rnwrite{Equation \noexpand\tag{#1} is a duplicate number.}\fi}%
  \csname @qnnum#1\endcsname\fi}

\def\ifunc@lled#1{\expandafter\ifx\csname @qnnum#1\endcsname\relax}

\let\@qnend=\end\gdef\end{\if@qnfile
\immediate\write16{Equation numbers written on []\jobname.EQN.}\fi\@qnend}

\catcode`@=12
\refstyleprnp
\catcode`@=11
\newcount\r@fcount \r@fcount=0
\def\refreset{\global\r@fcount=0}
\newcount\r@fcurr
\immediate\newwrite\reffile
\newif\ifr@ffile\r@ffilefalse
\def\w@rnwrite#1{\ifr@ffile\immediate\write\reffile{#1}\fi\message{#1}}

\def\writer@f#1>>{}
\def\referencefile{
  \r@ffiletrue\immediate\openout\reffile=\jobname.ref%
  \def\writer@f##1>>{\ifr@ffile\immediate\write\reffile%
    {\noexpand\refis{##1} = \csname r@fnum##1\endcsname = %
     \expandafter\expandafter\expandafter\strip@t\expandafter%
     \meaning\csname r@ftext\csname r@fnum##1\endcsname\endcsname}\fi}%
  \def\strip@t##1>>{}}

\def\citeall#1{\xdef#1##1{#1{\noexpand\cite{##1}}}}
\def\cite#1{\each@rg\citer@nge{#1}}	

\def\each@rg#1#2{{\let\thecsname=#1\expandafter\first@rg#2,\end,}}
\def\first@rg#1,{\thecsname{#1}\apply@rg}	
\def\apply@rg#1,{\ifx\end#1\let\next=\relax
\else,\thecsname{#1}\let\next=\apply@rg\fi\next}

\def\citer@nge#1{\citedor@nge#1-\end-}	
\def\citer@ngeat#1\end-{#1}
\def\citedor@nge#1-#2-{\ifx\end#2\r@featspace#1 
  \else\citel@@p{#1}{#2}\citer@ngeat\fi}	
\def\citel@@p#1#2{\ifnum#1>#2{\errmessage{Reference range #1-#2\space is bad.}%
    \errhelp{If you cite a series of references by the notation M-N, then M and
    N must be integers, and N must be greater than or equal to M.}}\else%
 {\count0=#1\count1=#2\advance\count1 by1\relax\expandafter\r@fcite\the\count0,
  \loop\advance\count0 by1\relax
    \ifnum\count0<\count1,\expandafter\r@fcite\the\count0,%
  \repeat}\fi}

\def\r@featspace#1#2 {\r@fcite#1#2,}	
\def\r@fcite#1,{\ifuncit@d{#1}
    \newr@f{#1}%
    \expandafter\gdef\csname r@ftext\number\r@fcount\endcsname%
                     {\message{Reference #1 to be supplied.}%
                      \writer@f#1>>#1 to be supplied.\par}%
 \fi%
 \csname r@fnum#1\endcsname}
\def\ifuncit@d#1{\expandafter\ifx\csname r@fnum#1\endcsname\relax}%
\def\newr@f#1{\global\advance\r@fcount by1%
    \expandafter\xdef\csname r@fnum#1\endcsname{\number\r@fcount}}

\let\r@fis=\refis			
\def\refis#1#2#3\par{\ifuncit@d{#1}
   \newr@f{#1}%
   \w@rnwrite{Reference #1=\number\r@fcount\space is not cited up to now.}\fi%
  \expandafter\gdef\csname r@ftext\csname r@fnum#1\endcsname\endcsname%
  {\writer@f#1>>#2#3\par}}

\def\ignoreuncited{
   \def\refis##1##2##3\par{\ifuncit@d{##1}%
     \else\expandafter\gdef\csname r@ftext\csname
r@fnum##1\endcsname\endcsname%
     {\writer@f##1>>##2##3\par}\fi}}

\def\r@ferr{\endreferences\errmessage{I was expecting to see
\noexpand\endreferences before now;  I have inserted it here.}}
\let\r@ferences=\references
\def\references{\r@ferences\def\endmode{\r@ferr\par\endgroup}}

\let\endr@ferences=\endreferences
\def\endreferences{\r@fcurr=0
  {\loop\ifnum\r@fcurr<\r@fcount
    \advance\r@fcurr by 1\relax\expandafter\r@fis\expandafter{\number\r@fcurr}%
    \csname r@ftext\number\r@fcurr\endcsname%
  \repeat}\gdef\r@ferr{}\global\r@fcount=0\endr@ferences}

\let\r@fend=\endpaper\gdef\endpaper{\ifr@ffile
\immediate\write16{Cross References written on []\jobname.REF.}\fi\r@fend}

\catcode`@=12

\citeall\refto		
\citeall\ref		%
\citeall\Ref		%

\referencefile

\def\ss{\scriptscriptstyle}

\def\frac#1/#2{#1 / #2}
\def\nugrant{This work was supported in part by the National Science
Foundation grants PHY-90-01439 and PHY-93-06906 (S.P.M.) and U.~S.~Department
of Energy grant DE-FG02-85ER40233 (M.T.V.)}
\def\neuphys{Department of Physics\\Northeastern University\\Boston MA 02115}

\def\oneandthreefifthsspace{\baselineskip=\normalbaselineskip
  \multiply\baselineskip by 8 \divide\baselineskip by 5}

\font\titlefont=cmr10 scaled\magstep3
\def\bigtitle                      
  {\null\vskip 3pt plus 0.2fill
   \beginlinemode \doublespace \raggedcenter \titlefont}

\def\msbar{\overline{\rm MS}}
\def\drbar{\overline{\rm DR}}
\def\mms{M_{\scriptscriptstyle{\overline{\rm MS}}}}
\def\mdr{M_{\scriptscriptstyle{\overline{\rm DR}}}}
\def\gms{g_{\scriptscriptstyle{\overline{\rm MS}}}}
\def\gdr{g_{\scriptscriptstyle{\overline{\rm DR}}}}
\def\ghat{{\hat g}_{\scriptscriptstyle{\overline{\rm MS}}}}
\def\bhat{{\hat b}} 
\oneandthreefifthsspace
\preprintno{NUB-3072-93TH}
\preprintno{August 1993}
\bigtitle{Regularization Dependence of Running Couplings
in Softly Broken Supersymmetry}
\author Stephen P. Martin and Michael T. Vaughn
\affil\neuphys
\body

\abstract
We discuss the dependence of running couplings on the choice of regularization
method in a general softly-broken $N=1$ supersymmetric theory. Regularization
by dimensional reduction respects supersymmetry, but standard dimensional
regularization does not. We find expressions for the differences between
running couplings in the modified minimal subtraction schemes of these two
regularization methods, to one loop order. We also find the two-loop
renormalization group equations for gaugino masses in both schemes,
and discuss the application of these results to the Minimal Supersymmetric
Standard Model.
\endtitlepage

\subhead{1. Introduction}
\taghead{1.}

Low-energy $N=1$ supersymmetry [\cite{reviews}]
provides an elegant solution to the naturalness problem associated with
the origin of the electroweak scale.
If supersymmetry proves to be correct, it is quite
possible that the first superpartner to be discovered will be the gluino,
because of its high production rate at hadron colliders.
Since the gluino mass is a physical parameter of the
supersymmetry-breaking sector, its determination will provide an
important clue as to the rest of the sparticle spectrum. Measurements of
the masses of the other superpartners will then yield tests of the various
specific extensions of the Minimal Supersymmetric Standard Model (MSSM) which
have been proposed over the years.
For example, in models with soft supersymmetry-breaking terms which are
``universal'' in the sense usually associated with spontaneously broken
supergravity [\cite{supergrav}],
the squark masses cannot be less than about $0.8$ of the gluino mass.
The details of the supersymmetric spectrum have been studied
under a variety of assumptions and constraints on the other unknown
parameters of the model
in [\cite{sp1},\cite{sp2},\cite{sp3}] among many others.
Further, there are sum rules [\cite{mrssc}] which relate, for example,
the masses of  the neutralinos and charginos to the mass of the gluino,
without involving  any other unknown input parameters.
It is therefore useful to work out
the predictions for sparticle masses in terms of the running
parameters of supersymmetric models, including radiative corrections.

There is a problem of principle, however, in discussing radiative corrections
in supersymmetric models. The most popular regularization scheme for discussing
radiative corrections within the Standard Model,
dimensional regularization [\cite{dreg}] (DREG), violates supersymmetry
explicitly because it introduces a mismatch between the numbers of gauge boson
and gaugino degrees of freedom. The modified scheme known as dimensional
reduction [\cite{dred}] (DRED) does not violate supersymmetry, and thus
maintains the supersymmetric Ward identities. In DREG, supersymmetry is
violated in the finite parts of one-loop graphs, and in the divergent parts
of two-loop graphs. This means that the $\beta$-functions for a supersymmetric
model will be different for the two schemes starting at the two-loop level.
Also, the running couplings computed in DREG with modified minimal
subtraction [\cite{msbar}] ($\msbar$) will differ from those computed in DRED
with modified minimal subtraction ($\drbar$) by virtue of finite one-loop
effects.

In this paper, we give the difference between running
couplings in the $\msbar$ and $\drbar$ schemes, to lowest non-trivial
order, for a general $N=1$ supersymmetric model with soft breaking. As
an application, we then consider the two-loop $\beta$-function for a
gaugino mass in both schemes. The $\msbar$ version
is obtained by  specializing the results of [\cite{MVI},\cite{MVII}] for
a general renormalizable theory to the case of a general $N=1$ supersymmetric
model. The two-loop $\beta$-function in the $\drbar$ scheme is then derived by
simply translating all $\msbar$ couplings into their $\drbar$ counterparts.
We do not enter here into the question of whether a fully consistent $\drbar$
calculation at the two-loop level actually exists; although some results
have been obtained for supersymmetric theories [\cite{tldred}],
the use of dimensional reduction in non-supersymmetric
theories is problematic. However, if such a consistent scheme exists, it
must reproduce the results obtained here to two-loop order.

We consider a general $N=1$ supersymmetric Yang-Mills
model. The chiral superfields $\Phi_i$ contain a complex scalar $\phi_i$
and a two-component fermion $\psi_i$ which transform as a (possibly reducible)
representation $R$ of the gauge group $G$. The superpotential is
$$
W = {1\over 6} Y^{ijk} \Phi_i \Phi_j \Phi_k
\eqno(superpotential)
$$
with $Y^{ijk} = (Y_{ijk})^*$.
In addition, the Lagrangian contains soft
supersymmetry-breaking terms of the form
$$
-{\cal L}_{SB} = {1\over 6} h^{ijk} \phi_i\phi_j\phi_k
+ {1\over 2}M \lambda \lambda + {\rm h.c.}
\eqno(soft)
$$
where $M$ is the mass of the gaugino $\lambda$.
Strictly speaking, the most general renormalizable softly-broken $N=1$
supersymmetric model also contains couplings in the superpotential
with dimensions of (mass) and (mass)$^2$, and soft supersymmetry-breaking
scalar couplings with dimensions of (mass)$^2$ and (mass)$^3$.
However, such terms are not relevant to the present discussion.

For simplicity we first give our results for the special case of a simple
[or $U(1)$] gauge group. We then explain the modifications required if the
gauge group is a direct product in Section 4, where we will also discuss
the MSSM.  We let ${\bf t^A} \equiv ({\bf t^A})_i^j$ denote the
representation matrices for the gauge group $G$.
Then
$$
({\bf t^A t^A})_i^j \equiv C(R) \delta_i^j
$$
$$
{\rm Tr}_R ({\bf t^A t^B}) \equiv  S(R) \delta^{AB}
$$
define the quadratic Casimir invariant $C(R)$ and the Dynkin index $S(R)$
for a representation $R$.
For the adjoint representation [of dimension denoted by $d(G)$],
$C(G) \delta^{AB} = f^{ACD} f^{BCD}$ with $f^{ABC}$ the
structure constants of the group.

\subhead{2. Dependence of Running Couplings on the Regularization Method}
\taghead{2.}

The strategy for determining the relationship between couplings
in the $\msbar$ and $\drbar$ schemes is to relate each running parameter
to a physical quantity which cannot depend on the choice of scheme.

First, consider the running gaugino mass parameter. By standard
techniques, one can compute the physical pole mass of the gaugino
to one-loop order as a function of the running mass parameter in
each scheme. In the $\msbar$ scheme, one finds
$$
M_{\rm pole} = \mms (\mu) \left [ 1 +
{\gms^2 \over 16 \pi^2 } \left \lbrace
4 C(G) + 3 C(G) {\rm ln}  {\mu^2\over \mms^2  }
+ \sum_{\Phi} A_\Phi \right \rbrace
 \right ]\>,
\eqno(mgmsbar)
$$
while in the $\drbar$ scheme, the result is
$$
M_{\rm pole} = \mdr(\mu) \left [ 1 + {\gdr^2 \over 16 \pi^2 }
\left \lbrace
5 C(G) + 3 C(G) {\rm ln}{\mu^2\over \mdr^2 }
+ \sum_\Phi A_\Phi \right\rbrace
 \right ]\>.
\eqno(mgdrbar)
$$
In both cases
$$
A_\Phi \equiv 2  S(R_\Phi) \int^1_0 dx \> x\>
{\rm ln}\left ( [x m_\phi^2 + (1-x) m_\psi^2 - x(1-x) M_{\rm pole}^2 ]
/ \mu^2 \right )
$$
and the sum is over all chiral supermultiplets $\Phi = (\phi,\psi)$
which couple to the gaugino. Comparing \(mgmsbar) and \(mgdrbar), we obtain
(to one-loop order)
$$
\mms = \mdr \left [
1 + {g^2\over 16 \pi^2} {C(G)} \right ]
\>\> .
\eqno(mgmsdr)
$$

Similarly, we can compute the relationship between the gauge couplings
in the two schemes. Here we need to make an important distinction
between the gauge coupling $g$ which appears in the interactions of
the vector gauge bosons, and the coupling $\hat g$ which occurs in the
Yukawa interaction
$$
{\cal L}_{\hat g} = {\sqrt 2} {\hat g} \phi^i ({\bf t^A})_i^j (\psi_j
\lambda^A)
+ {\rm H.~c.}
$$
of the gaugino and the chiral superpartners.
Gauge invariance guarantees that $g$ is the same everywhere
it appears, but only supersymmetry guarantees that $\hat g = g$. Hence
we have $\hat g = g$ in DRED, but we expect
$\ghat \not= \gms$ in DREG, since the radiative corrections in DREG violate
supersymmetry. By computing the relevant one-loop
graphs in the effective action, and demanding that the physical scattering
amplitudes computed in each scheme are the same, we find
$$
\gms = \gdr \left [
1 - {g^2\over 96 \pi^2} C(G) \right ]
\eqno(gmsbar)
$$
and
$$
\ghat  = \gdr \left\{
1 +  {g^2\over 32 \pi^2} \left[C(G)- C(r)\right]\right\} \>\> .
\eqno(ghatmsbar)
$$
Thus there is a different coupling $\ghat$ of the gaugino to each
distinct irreducible representation $r$ of the chiral supermultiplets.
Note that we consistently neglect the distinction between
$\gms$, $\ghat$, and $\gdr$ in the one-loop correction parts
of \(mgmsdr), \(gmsbar), and \(ghatmsbar), and in the two-loop correction
parts of formulas below, since all of the incarnations of the gauge coupling
are the same to zeroth order. The relations \(mgmsdr) and \(gmsbar) have
been discussed by Yamada [\cite{yamada}] (see also [\cite{AKT},\cite{SSK}])
for a theory with only gauge
vector supermultiplets. Here we note that these formulas are not modified
by the presence of chiral superfields, which is not
surprising  in view of the fact that DRED and DREG only differ for graphs
in which there is at least one internal gauge boson line which does not
terminate (at either end) in a scalar-scalar-gauge boson vertex.

As a non-trivial consistency check, we consider the two-loop
running of both $\gms$ and $\ghat$, by specializing
the results of [\cite{MVI},\cite{MVII}] for a general renormalizable
theory to the case of $N=1$ supersymmetry. The
gauge coupling satisfies
$$
\eqalignno{
& \mu {d\over d\mu} \gms =  \> {1\over 16 \pi^2} b^{\ss (1)}
+ {1\over (16 \pi^2)^2 } b^{\ss (2)} &(betag)
\cr
b^{\ss (1)} = & g^3 \left[S(R) - 3 C(G) \right]
\cr
b^{\ss (2)} = & g^5 \left\{ - 6[C(G)]^2 + 2 C(G) S(R) + 4 S(R) C(R) \right\}
    - g^3 Y^{ijk} Y_{ijk}C(k)/d(G)
}
$$
(with all $\msbar$ couplings on the RHS)
as is already known [\cite{known}]. Here $S(R)$ is the Dynkin index summed
over all chiral multiplets and $S(R)C(R)$ is the sum of the Dynkin indices
weighted by the quadratic Casimir
invariant. Note that the gauge coupling $\gdr$ also satisfies eq.~\(betag) with
$\gms$ replaced everywhere by $\gdr$ according to \(gmsbar);
this is the well-known result that the $\beta$-function for the gauge
coupling is scheme-independent through two loops.

On the other hand, for the coupling of a gaugino to a chiral
supermultiplet in the irreducible representation $r$, we find from
[\cite{MVII}]
$$
\eqalignno{
& \mu {d\over d\mu} \ghat =  {1\over 16 \pi^2}
\bhat^{\ss (1)}+ {1\over (16 \pi^2)^2 } \bhat^{\ss (2)} &(betaghat)
\cr
\bhat^{\ss (1)} = & \ghat \left[\ghat^2 S(R) - 3 \gms^2 C(G)
+ 3 (\ghat^2- \gms^2) C(r)\right]
\cr
\bhat^{\ss (2)} = &  g^5 \left\{ -10 [C(G)]^2 + 2 C(G) S(R) + 5 S(R) C(R)
- S(R) C(r) - C(G) C(r) + 3 [C(r)]^2 \right\}\cr
 &\hbox{\hskip 160pt}- g^3 Y^{ijk} Y_{ijk}C(k)/d(G) \cr
}
$$
Here $\ghat^2 S(R)$ means a sum over all chiral multiplets of the Dynkin index
times the appropriate $\ghat^2$.
It is easy to check that \(betaghat)
is indeed consistent with \(gmsbar), \(ghatmsbar), and \(betag).

We now consider the Yukawa coupling  $Y^{ijk}$
between a scalar $\phi_i$ and two chiral fermions $\psi_j, \psi_k$.
By  again demanding that
physical scattering amplitudes computed from the one-loop effective
action in each scheme are the same, we find
$$
\left [Y_{\ss \msbar}^{i} \right ]^{jk}
= Y^{ijk}_{\ss \drbar} \left\{1 + {g^2\over 32 \pi^2}[C(r_j) + C(r_k)
- 2 C(r_i)]\right\}
\>\> .\eqno(ymsbar)
$$
Note that the $Y^{ijk}$ are totally symmetric in the
supersymmetry-respecting $\drbar$ scheme
(because of the way they appear in the superpotential), but not in
the $\msbar$ scheme due to the radiative corrections.
In $N=1$ supersymmetry, the scalar quartic interaction
$$
{\cal L}_{\rm quartic} = -{1\over 4} {\lambda_{ij}}^{kl} \phi^{*i} \phi^{*j}
\phi_k \phi_l
\eqno(quartic)
$$
depends entirely on the Yukawa and gauge couplings. In a
supersymmetry-respecting scheme like $\drbar$, this dependence is given
simply by
$$
{\lambda_{ij}}^{kl} = Y_{ijm}Y^{klm} + g^2 \left (
{\bf t^A}_i^k {\bf t^A}_j^l + {\bf t^A}_j^k {\bf t^A}_i^l \right )\>\> .
\eqno(drbarqu)
$$
However, in $\msbar$ this relation is modified by radiative corrections.
Indeed, we find
$$
{\left [\lambda_{\ss \drbar} \right ]_{ij}}^{kl}
-{\left [\lambda_{\ss \msbar} \right ]_{ij}}^{kl}
= {g^4 \over 16 \pi^2}
\lbrace {\bf t^A , t^B} \rbrace_i^k
\lbrace {\bf t^A , t^B}  \rbrace_j^l
+ (i \leftrightarrow j)  \> ,
\eqno(quarcorr)
$$
again by comparing physical scattering amplitudes derived from the
one-loop effective actions in the two schemes. This completes the
``dictionary'' for translating between $\msbar$ and $\drbar$ couplings
at one loop order. A gauge-invariant, supersymmetric fermion mass
coming from a quadratic term in the superpotential will also differ
between the two schemes, but that difference is trivially derived
from \(ymsbar) by taking the scalar with index $i$ to be a dummy field
with
$C(r_i)=0$ and $C(r_j)= C(r_k)$.
None of the other couplings
(written in component language rather than superfield language)
of a general softly-broken $N=1$ supersymmetric model
differ between $\msbar$ and $\drbar$ to this order, because
DREG and DRED only differ for graphs containing a gauge boson line
which does not end in a scalar-scalar-gauge boson vertex.

\subhead{3. Two-Loop Running of Gaugino Masses}
\taghead{3.}

In this section we will consider the two-loop $\beta$-functions for the
gaugino mass computed in both DREG and DRED.
The results of [\cite{MVII}] provide the 2-loop $\msbar$ $\beta$-function for
a scalar-fermion-fermion coupling in a general renormalizable theory. From
this, it is trivial to extract the corresponding result for
a fermion mass term, by treating the external scalar in eqs.~(3.3) and (3.4)
of [\cite{MVII}] as a dummy field
(with no gauge or other Yukawa interactions). After further specializing
to the case of a gaugino mass in $N=1$ supersymmetry,
and using \(gmsbar) and \(ghatmsbar) to eliminate $\ghat$ in favor of $\gms$,
we obtain:
$$
\eqalignno{
&\mu {d\over d\mu} \mms = {1\over 16 \pi^2} \gamma^{\ss (1)}_{\ss \msbar}
+ {1\over (16 \pi^2)^2 } \gamma^{\ss (2)}_{\ss \msbar}
&(betammsbar)
\cr
\gamma^{\ss (1)}_{\ss \msbar}
= & \gms^2 \left[ 2 S(R) - 6 C(G) \right] \mms
\cr
\gamma^{\ss (2)}_{\ss \msbar}
= & g^4\left\{ -32[ C(G)]^2 +{\textstyle{32\over 3}}C(G)S(R)
+ 16 S(R)C(R)\right\}M\cr
 &\hbox{\hskip 160pt} + 2 g^2 \left [h_{ijk}  - M Y_{ijk}\right] Y^{ijk}
C(k)/d(G) \>\> .
\cr
}
$$

The simplest way to find the corresponding equation in DRED
is to plug eqs.~\(mgmsdr) and \(gmsbar) into \(betammsbar).  This gives
$$
\eqalignno{
& \mu {d\over d\mu} \mdr = {1\over 16 \pi^2} \gamma^{\ss (1)}_{\ss \drbar}
+ {1\over (16 \pi^2)^2 } \gamma^{\ss (2)}_{\ss \drbar} &(betamdrbar)
\cr
\gamma^{\ss (1)}_{\ss \drbar}
= & \gdr^2 \left[ 2 S(R) - 6 C(G) \right] \mdr
\cr
\gamma^{\ss (2)}_{\ss \drbar}
= & g^4\left\{ -24[C(G)]^2 + 8 C(G) S(R) + 16  S(R)C(R)\right\} M\cr
 &\hbox{\hskip 160pt} + 2 g^2 \left [h_{ijk}  - M Y_{ijk}\right] Y^{ijk}
C(k)/d(G) \>\> .
\cr
}
$$
The method used here avoids potential ambiguities and conceptual
problems [\cite{tldred}] associated with doing
two-loop calculations directly in DRED. We will return to these
issues elsewhere. Eqs.~\(betammsbar) and \(betamdrbar) generalize the
expressions given by Yamada[\cite{yamada}], who pointed out that the one-loop
relation $\beta_M = (2M/g) \beta_g $ fails to hold at the two-loop level.

\subhead{4. Direct Product Gauge Groups and the MSSM}
\taghead{4.}

As promised, we now point out the modifications which
must be made to the preceding formulas if the gauge group is a product of
simple [or $U(1)$] subgroups $G_a$. The formulas
involving radiative corrections for a gauge coupling
$g_a$ [\(gmsbar)-\(betaghat)] or gaugino mass $M_a$ [\(mgmsbar)-\(mgmsdr)
and \(betammsbar)-\(betamdrbar)]
for each subgroup $G_a$ may be obtained by the following set of
rules [\cite{MVI}]. First, each term which does not involve the quadratic
Casimir invariant of a non-adjoint representation is diagonal in subgroups;
i.e., obtained by $g\rightarrow g_a$, $S(R)\rightarrow S_a(R)$, and
$C(G) \rightarrow C(G_a)$. For the other terms, one has
$ g^2 C(r) \rightarrow \sum_b g_b^2 C_b(r) $
in \(ghatmsbar) and \(ymsbar);
$C(k)/d(G) \rightarrow C_a(k)/d(G_a)$ in \(betag), \(betaghat),
\(betammsbar), and \(betamdrbar);
and
$$
g^5 S(R) C(R)  \rightarrow g_a^3 S_a(R) \sum_b g_b^2 C_b(R)
$$
in eq.~\(betag).
[Consistency then determines the replacement rules for the
$C(r)$-dependent terms in \(betaghat).] The correction term in \(quarcorr)
becomes a double sum $\sum_b \sum_c g_b^2 g_c^2 \ldots $
over subgroups. Finally, in eqs.~\(betammsbar) and \(betamdrbar) we need
the replacement
$$
16 g^4 C(R) S(R) M \rightarrow 8 g_a^2 S_a(R) \sum_b g_b^2 C_b(R) (M_a + M_b)
\>\> .
$$

In the MSSM, the gauge group is $SU(3)_c\times SU(2)_{\ss L}
\times U(1)_{\ss Y}$, with chiral superfields $Q$ and $L$ for the
$SU(2)_{\ss L}$-doublet quarks and leptons, and $u$, $d$, $e$ for the
$SU(2)_{\ss L}$-singlet quarks and leptons, and two Higgs doublet chiral
superfields $H_u$ and $H_d$. The relevant part of the superpotential is
$$
W = H_u Q Y_u u + H_d Q Y_d d + H_d L Y_e e
\eqno(mssmsuper)
$$
where $Y_u$, $Y_d$, $Y_e$ are $3\times 3$ Yukawa matrices. The
soft supersymmetry-breaking Lagrangian includes trilinear scalar couplings
$$
-{\cal L} = H_u Q h_u u + H_d Q h_d d + H_d L h_e e + {\rm h.c.}
\eqno(mssmsoft)
$$
where $h_{u,d,e}$ are again $3\times 3$ matrices in family space, and
we use the same symbol for scalar components as for the chiral superfield.
Then in the $\drbar$ scheme the 2-loop running of the three gauge couplings
are given by
$$
\mu {d\over d\mu} g_a =  {g_a^3\over 16\pi^2} B^{\ss (1)}_a +
{g_a^3\over (16 \pi^2)^2} \left [  B_a^{\ss (2)} g_a^2 +
\sum_{b=1}^3 B^{\ss (2)}_{ab} g_b^2 -
\sum_{x=u,d,e} C_a^x \, {\rm Tr }   Y_xY_x^\dagger \right ]
\>\> .
\eqno(mssmg)
$$
Here $B_a^{\ss (1)} = (33/5,1,-3)$ for $U(1)_{\ss Y}$ (in a GUT normalization),
$SU(2)_{\ss L}$, and  $SU(3)_c$ respectively; $B_a^{\ss (2)} = (0,4,-18)$, and
$$
B^{\ss (2)}_{ab} = \pmatrix{ {199/ 25} & {27/5} & {88/ 5} \cr
                 {9/ 5}    & 21      & 24          \cr
                 {11/ 5}   &  9          & 32          \cr}
\qquad {\rm and} \qquad
C^{u,d,e}_a = \pmatrix{ {26/ 5} & {14/ 5} & {18/ 5} \cr
                          6         & 6           & 2           \cr
                          4         & 4           & 0           \cr }
\>\> .
$$
The renormalization group equation for the three gaugino mass parameters
in $\drbar$ can then be written easily in terms of the same coefficients:
$$
\eqalign{
\mu {d\over d\mu} M_a = & {2 g_a^2\over 16\pi^2} B^{\ss (1)}_a M_a +
{2 g_a^2\over (16 \pi^2)^2} \biggl [ 2  B_a^{\ss (2)} g_a^2 M_a
\cr & + \sum_{b=1}^3 B^{\ss (2)}_{ab} g_b^2 (M_a + M_b)
+ \sum_{x=u,d,e} C_a^x \left ( {\rm Tr} Y_x h_x^\dagger -
M_a {\rm Tr } Y_xY_x^\dagger \right ) \biggr ]
\cr }
\eqno(smgm)
$$

In the $\msbar$ scheme, the two-loop renormalization group equations
for the gauge couplings are exactly the same as in $\drbar$ (provided,
of course, that one consistently uses $\gms$ rather than $\ghat$). The $\msbar$
two-loop renormalization group equations for the gaugino mass parameters in
the MSSM are given by an equation exactly like \(smgm), with the only
difference that $B_a^{\ss (2)}$  must be  replaced in \(smgm) by
$B_a^{\ss (2) \msbar} = (0,16/3, -24)$.

\subhead{5. Conclusion}
\taghead{5.}

In this paper, we have presented analytical expressions for the differences
between couplings in the $\msbar$ and $\drbar$ schemes, to one loop
order. The $\msbar$ scheme,
while convenient for analyzing the predictions of the standard model,
is slightly inconvenient for the MSSM because it violates supersymmetry,
in the sense that the tree-level supersymmetric relations between coupling
constants are modified by finite radiative corrections even at one loop.
We were able to exploit this fact to derive the two-loop $\beta$-function
for a gaugino mass in $\drbar$ from knowledge of the corresponding
formula in the $\msbar$ scheme, which in turn followed from
the results for a general renormalizable theory  (using $\msbar$)
given in [\cite{MVI},\cite{MVII}].
The same method can in fact be used to derive the
renormalization group equations for all of the other soft
supersymmetry-breaking parameters of the MSSM. We will report on
this in a future publication.

Let us close by remarking on the significance of our results for predictions
in supergravity-inspired models. These are typically obtained by choosing
a set of input parameters, including a common gaugino mass, at some very
high scale and running the parameters down to the electroweak
scale. The boundary conditions at the very high scale should presumably
be applied in a supersymmetry-respecting scheme like $\drbar$, and
threshold effects are simpler [\cite{AKT}] in $\drbar$. However, it may
be more convenient at low energies to work in $\msbar$.
In the MSSM, the two-loop contributions can lower the predicted value of
the gluino mass by several percent compared to the one-loop predictions, while
having a much smaller effect on the other gaugino masses.  (In extensions
of the MSSM with non-minimal particle content, these corrections are
potentially much larger.) Another effect which
is at least as important is the proper treatment of the gluino pole mass,
as given by eqs.~\(mgmsbar) and \(mgdrbar). One might choose
to evaluate the running mass at a scale equal to the running mass at that
scale (i.e., solve the equation $\mms (\mu) = \mu$ or $\mdr (\mu) = \mu$,
depending on which scheme one is working in);
however, this will always underestimate the true pole mass of the gluino,
and often by more than 5 per cent. To roughly estimate the size of the effect,
suppose the squarks are taken to be degenerate with the gluino and much
heavier than all quarks; then from \(mgdrbar) one finds
$M_{\rm pole} \approx \mdr(\mdr ) [ 1+ 9 \alpha_3/4 \pi ]$.
Such considerations may have an effect
on future efforts to understand the parameters of the MSSM.

S.P.M. thanks Diego Casta\~no and David G.~Robertson for helpful
conversations, and the Physics Department of Southern Methodist University
for their warm hospitality.
M.T.V. thanks the Physics Department and the High Energy Theory group at
the University of Southampton for their hospitality during the completion
of this work.
\nugrant

\references

\refis{dreg} G.~'t Hooft and M.~Veltman, \np B44, 189, 1972.

\refis{tldred}
I.~Jack, \pl 147B, 405, 1984; I.~Jack and H.~Osborn, \np B249, 472, 1985;
R.~van Damme and G.~'t Hooft, \pl 1550B, 133, 1985.

\refis{dred} W.~Siegel, \pl 84B, 193, 1979; D.~M.~Capper, D.~R.~T.~Jones
and P.~van~Nieuwenhuizen, \np B167, 479, 1980.

\refis{msbar} W.~A.~Bardeen, A.~J.~Buras, D.~W.~Duke and T.~W.~Muta,
\pr D18, 3998, 1978.

\refis{MVI} M.~E.~Machacek and M.~T.~Vaughn,
\np B222, 83, 1983.

\refis{MVII} M.~E.~Machacek and M.~T.~Vaughn,
\np B236, 221, 1984.

\refis{known} D.~R.~T.~Jones, \np B87, 127, 1975;
D.~R.~T.~Jones and L.~Mezincescu, \pl 136B, 242, 1984.

\refis{mrssc} S.~P.~Martin and P.~Ramond, ``Sparticle Spectrum Constraints'',
Northeastern University preprint NUB-3067-93TH, June 1993.

\refis{reviews}
For reviews, see H.~P.~Nilles,  \prpts 110, 1, 1984 or
H.~E.~Haber and G.~L.~Kane, \prpts 117, 75, 1985.

\refis{supergrav}
A.~Chamseddine, R.~Arnowitt and P.~Nath,
\journal Phys.~Rev.~Lett., 49, 970, 1982;
H.~P.~Nilles,
\journal Phys.~Lett., 115B, 193, 1982;
L.~E.~Ib\'a\~nez,
\journal Phys.~Lett., 118B, 73, 1982;
R.~Barbieri, S.~Ferrara, and C.~Savoy,
\journal Phys.~Lett., 119B, 343, 1982;
L. Hall, J. Lykken and S. Weinberg,
\journal Phys.~Rev., D27, 2359, 1983;
P. Nath, R. Arnowitt and A. H. Chamseddine,
\journal Nucl.~Phys., B227, 121, 1983.

\refis{AKT} I.~Antoniadis, C.~Kounnas, and K.~Tamvakis,
\journal Phys. Lett., 119B, 377, 1982.

\refis{SSK} G.~A.~Schuler, S.~Sakakibara, and J.~G.~K\"orner,
\journal Phys. Lett., 194B, 125, 1987.

\refis{sp1}
G.~G.~Ross and R.~G.~Roberts,
\journal Nucl.~Phys., B377, 571, 1991;
L.~E.~Ib\'a\~nez and  G.~G.~Ross,
``Electroweak Breaking in Supersymmetric Models'',
CERN-TH.6412/92, in Perspectives in Higgs Physics,
G.~Kane, editor.

\refis{sp2}
R. Arnowitt and P. Nath,
SSCL-Preprint-229;
\prl 69, 725, 1992;
\pr D46, 3981, 1992; P.~Nath and R.~Arnowitt,
\pl B289, 368, 1992; ibid, {\bf B287} (1992) 89.

\refis{sp3}
P.~Ramond, ``Renormalization Group Study of the Minimal Supersymmetric
Standard Model: No Scale Models'', invited talk at workshop ``Recent
Advances in the Superworld'', Houston TX, April 1993;
D.~J.~Casta\~no, E.~J.~Piard, and P.~Ramond, University of Florida
preprint UFIFT-HEP-93-18.

\refis{yamada} Y.~Yamada, ``Two-loop renormalization of gaugino mass
in supersymmetric gauge model'', Tokyo preprint UT-647, June 1993.

\endreferences\end